\documentclass[aps,pra]{revtex4-2}

\usepackage{amsmath,amssymb}
\usepackage{graphicx}
\usepackage{hyperref}
\usepackage[dvipsnames,svgnames,x11names]{xcolor}
\usepackage{amsmath,amssymb,amsfonts}
\usepackage{mathtools}
\usepackage{float}
\usepackage{setspace}

\begin{document}

\title{Semiclassical Quantum Mechanics and the Optics of Time-Varying Media}
\author{S. A. R. Horsley}
\affiliation{School of Physics and Astronomy, University of Exeter, Stocker Road, Exeter, EX4 4QL, UK}
\author{R. Sokol}
\affiliation{Faculty of Science, Palacký University Olomouc, 17. listopadu 1192/12, 779 00 Olomouc, Czech Republic}
\author{E.-M. Graefe}
\affiliation{Department of Mathematics, Imperial College London, South Kensington, London, SW7 2AZ}

\begin{abstract}
Rapid, high contrast modulation of optical parameters such as the refractive index can radically reshape a wave's frequency spectrum as it propagates.  Non-commuting frequency and time operators can be used to describe this reshaping, analogous to the position and momentum operators of quantum mechanics.
Here we develop the connection between this theory and semi-classical quantum mechanics, showing that: (i) different microscopic models of time-varying materials are equivalent to choices of operator ordering; (ii) wave propagation is governed by an unusual non-Hermitian Schr\"odinger equation; (iii) we can take a semi-classical limit, finding a generalization of Hamilton's equations governing the central frequency and arrival time of a wave spectrum as it propagates through a modulated material.  The theory can be applied to design dispersive time modulations that `focus' a wave in frequency and time, analogous to caustics in semiclassical quantum mechanics.  Our results not only deepen the connection between classical wave propagation and quantum mechanics, but can also simplify overly complex calculations that have been previously applied to media where time reflection effects are negligible. 
\end{abstract}
\maketitle
\section{Introduction}
Quantum mechanics grew out of the realisation that the conjugate variables of classical mechanics are an approximation to a deeper, wave theory.  Position and momentum variables are promoted to operators that do not commute, which implies that no state of a system can be  a simultaneous eigenfunction of both, i.e., there is an uncertainty relation between position and momentum.  Although usually introduced as a hallmark of quantum mechanics, this operator description is not peculiar to quantum theory.  It is largely a historical accident that other theories are not framed in this way.  For example, the vacuum Maxwell equations become a massless spin-1 Dirac equation when the spatial derivatives are rewritten as components of the momentum operator, a result that has been used to study optical angular momentum~\cite{Barnett2014}, electromagnetic interface states~\cite{Horsley2018,Horsley2019,Bliokh2019}, and develop an analogy between Maxwell's and Dirac's theories~\cite{Bialynicki1996,Dennis2023}.  Not only does the use of the operator language make the connection between quantum theory and other wave phenomena easier to understand, but it also enables the application of different mathematical machinery to problems in classical wave-propagation.  This paper is a continuation of this idea, where we apply the formalism of semi-classical quantum mechanics to study wave propagation in dispersive, time-varying materials.

Time, or space-time varying media are the subject of a rapidly growing research field in classical wave physics, where---for electromagnetic waves---the permittivity and/or permeability, are explicit functions of time, breaking temporal translation symmetry and allowing energy to be pumped into, or out of the field.  This time dependence is nearly always due to some external source of energy, be that an acceleration of the laboratory~\cite{bacot2016}, an intense optical ``pump'' field~\cite{Zhou2020,Bohn2021,Tirido2023,Lustig2023}, or electronic modulation~\cite{Moussa2023,Hooper2025,Wen2022}.  Theoretical work on this subject can be traced back to Morgenthaler~\cite{morgenthaler1958} and later studies of wave propagation in driven plasmas~\cite{mendona2002}.  The revival of interest has largely arisen from experiments making use of both metamaterial concepts, and the discovery of materials where the permittivity can be changed on the order of unity, over tens of femtoseconds (see Ref.~\cite{galiffi2022} for a review).  Such rapid, high contrast modulation of the material parameters enables effects such as nonreciprocity~\cite{sounas2017}, synthetic motion~\cite{harwood2025}, and amplification, and leads to phenomena such as temporal refraction and reflection~\cite{bacot2016,Moussa2023,Lustig2023} and the formation of bandgaps in momentum rather than frequency~\cite{lustig2023b}.  More recent work has connected these systems to non-Hermitian physics and exceptional points~\cite{koutserimpas2020,Hooper2025}, which we also do here.

\begin{figure}[h!]
\begin{center}
\includegraphics[width=\textwidth]{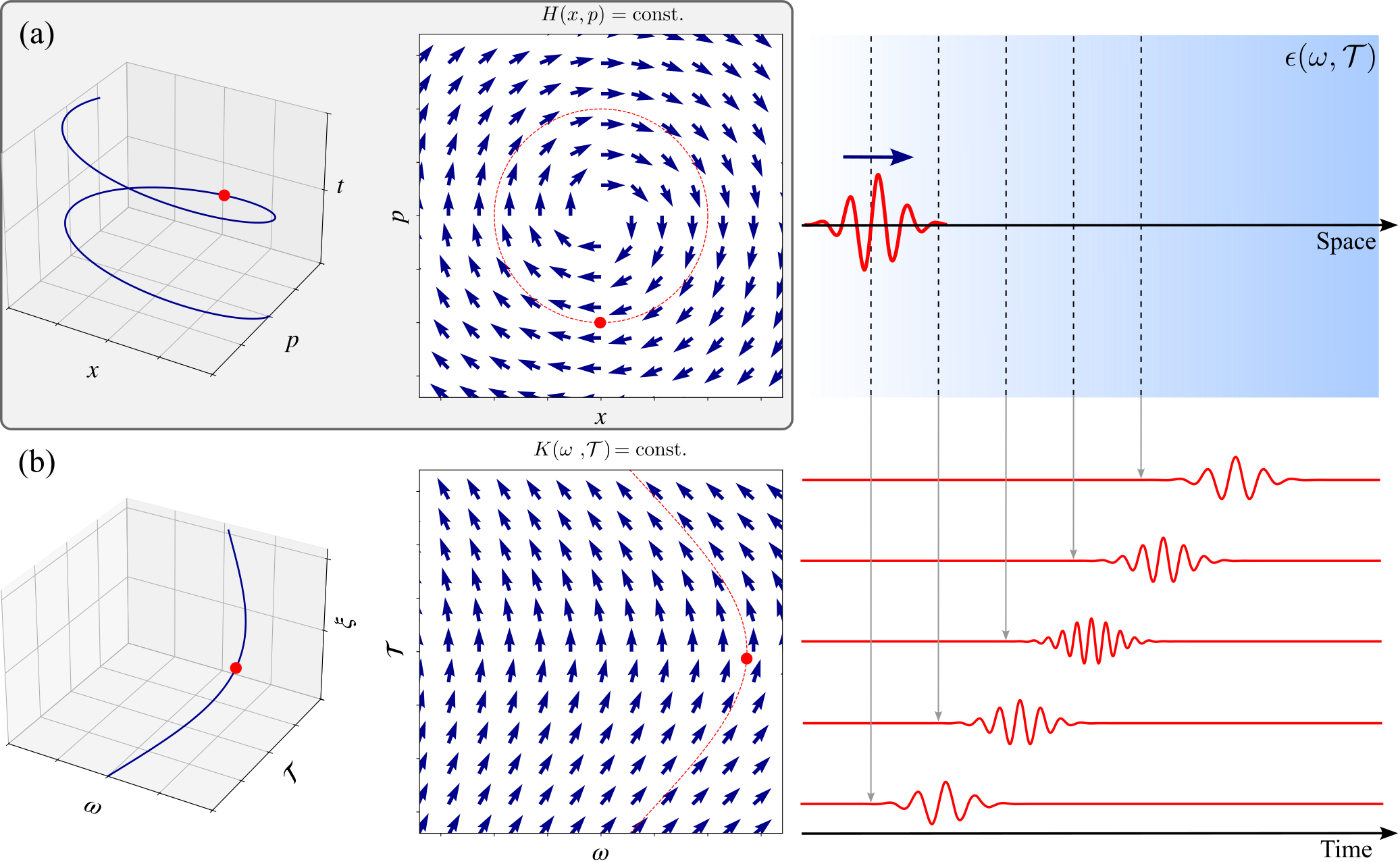}
\end{center}
\caption{\textbf{Phase space and time-varying media:} \textbf{(a)} Classical mechanics describes the trajectory of a particle as a point (here red, showing harmonic motion) in position and momentum coordinates, known as \emph{phase space}.  The Hamiltonian $H(x,p)$ governs this motion, with $\dot{p}=-\partial H/\partial x$ and $\dot{x}=\partial H/\partial p$ generating a `flow' in phase space (blue vector field). The motion of a particle follows this flow, which conserves energy, ${H}(x,p)=\text{const.}$.  Quantum mechanics replaces the phase-space coordinates with non-commuting operators $[\hat{x},\hat{p}]={\rm i}\hbar$. When the action $S$ is large compared to $\hbar$, non-commutativity is negligible and classical mechanics is recovered.  \textbf{(b)}  A similar phase-space picture can be used to understand wave propagation in time-varying media.  We can specify the state of a wavepacket at each spatial coordinate $\xi$ in a material in terms of its central frequency $\omega$ and average time of arrival $\mathcal{T}$.  As shown on the right, we imagine measuring the time evolution of the wavepacket at successive depths into a homogeneous medium, observing the shift in arrival time and central frequency, as indicated by the trajectory shown in the left hand plot.  Due to the spatial translational symmetry of the material, the phase-space flow is---in the Hermitian case---governed by the conservation of wave-vector $K(\omega,\mathcal{T})=\text{const.}$, which generates an analogous phase-space flow.  As in the case of classical and quantum mechanics,  frequency and momentum are better described by non-commuting operators $[\hat{\omega},\hat{\mathcal{T}}]={\rm i}\bar{h}$, where time $\mathcal{T}$ is a measured in units of the (slow) modulation timescale $T$, and frequency $\omega$ in units of $\Delta^{-1}$ where $\Delta$ is the (fast) timescale of the material response, and we have defined the small \textit{semiclassical} parameter $\bar{h}=\frac{\Delta}{T}$. When the time-modulation of the medium is slow, $\bar{h}$ is small, and we recover a description of the wave propagation with commuting phase-space coordinates $\omega$ and $\mathcal{T}$ in a semiclassical limit.}\label{fig:phase-space}
\end{figure}

Here we take the perspective that the conjugate variables describing wave propagation in a time varying medium are the \textit{frequency} $\omega$ of a propagating pulse and the central arrival \textit{time} $\mathcal{T}$ observed at a fixed point in space~\cite{Horsley2023}.  These can be understood as phase-space coordinates, analogous to position and momentum (see Fig.~\ref{fig:phase-space}).  Unlike classical mechanics, where points in phase space evolve as a function of time, points in $(\omega,\mathcal{T})$ space evolve as a function of propagation distance, $\xi$, changing as we move to different points within the material.  For example, for propagation in free space, the trajectory in phase space is simply a point of constant frequency that moves to ever larger times with increasing $\xi$.  In a time-varying-dispersive material, the trajectories can be much richer.


%
%
As in classical mechanics, the phase-space trajectory is only an approximation.  The exact wave dynamics is obtained through promoting the phase-space coordinates to non-commuting operators.  The situation is no different here, where frequency and time become operators.  However, as mentioned above, one important difference is that -- unlike \textcolor{magenta}{in} quantum mechanics -- the operator governing the evolution of the wave in a time-varying medium is nearly always non-Hermitian. There are several physical reasons for this. The most obvious is the presence of dissipation in wave propagation. Unlike the net probability contained in a quantum wave, the energy contained in a classical electromagnetic wave is typically not conserved during propagation, being lost into the dynamics of the material. Thus, in a sense, classical wave physics provides a richer dynamics than quantum theory.  In the last fifteen or so years this connection has motivated an experimental investigation of PT-symmetric Hamiltonians~\cite{Makris2008,Longhi2010,ElGanainy2019,Ruter2010,Guo2009} (which were previously a hypothetical extension of quantum mechanics~\cite{Bender1998,Bender2024}), in addition to probing the peculiar dynamics in systems close to a non-Hermitian degeneracy (an ``exceptional point'')~\cite{Berry2011,Lee2009,Miri2019}, proposed as a route to enhanced sensors~\cite{Wiersig2016}.  Here we show that time-varying media provide a platform where this connection between non-Hermitian quantum theory and classical wave propagation is perhaps even deeper.

%
%
%
\section{Constitutive relations and the `quantization' of phase space\label{sec:ordering}}
%
%
%

A typical, unmodulated medium is characterized in terms of a dielectric susceptibility $\chi$ that is a function only of the difference between the observation time, and the time when the polarizability is measured.  The susceptibility of a time-varying medium depends on \emph{both} the time since the electric field was applied, \emph{and} the time of observation $t$.  The constitutive relations are often expressed as
\begin{equation}
    \boldsymbol{P}(t)=\epsilon_0\int_{-\infty}^{t}{\rm d}t'\,\chi(t,t-t')\boldsymbol{E}(t').
\end{equation}
However, as discussed in e.g.~\cite{HorsleyBaker2025,Dixon2026} writing the susceptibility as an explicit function of $t$ rather than $t'$ is a choice that assumes a particular model of the material dynamics.  Here we write, more generally
\begin{equation}
 \boldsymbol{P}\left(t\right)=\epsilon_0\int_{-\infty}^{t}{\rm d}t'\,\chi\left(\alpha t+(1-\alpha)t',t-t'\right)\boldsymbol{E}\left(t'\right),\label{eq:polarization-alpha}
\end{equation}
where we have introduced the parameter $\alpha$ to specify different models for the dynamics of the medium, the time modulation either occurring at the observation time $t$ ($\alpha=1$), the time when the field was applied $t'$ ($\alpha=0$), or somewhere in between (e.g. $\alpha=1/2$ for the mid-point).
\\
We now show that this model dependence of the material dynamics in Eq. (\ref{eq:polarization-alpha}) is equivalent to a standard quantisation procedure of a classical phase-space function allowing for different orderings of the position and momentum operators.  To show this we first transform Eq. (\ref{eq:polarization-alpha}) into the Fourier domain.  After a change of variables and carrying out a Fourier integral (frequencies $w$ and $w'$), Eq. (\ref{eq:polarization-alpha}) becomes
\begin{equation}
\tilde{\boldsymbol{P}}\left(w\right)=\epsilon_0\int_{-\infty}^{\infty}\frac{{\rm d}w'}{2\pi}\int_{-\infty}^{\infty}{\rm d}t\,{\rm e}^{{\rm i}(w-w')t}\tilde{\chi}\left(t,\alpha w'+(1-\alpha)w\right)\tilde{\boldsymbol{E}}\left(w'\right)\label{eq:intermediate},
\end{equation} 
where `\,$\tilde{\;}$\,' indicates a quantity evaluated in the Fourier domain, with $\tilde{\chi}$ the Fourier transform of $\chi$ with respect to the second argument. 
 Now introducing two time-scales, the `fast' time-scale $\Delta$ which represents the response time of the medium (assumed comparable to the period of oscillation of the field), and the `slow' timescale $T$, which is the time over which the material response is modulated, and rescaling frequency and time variables, introducing  $\omega=w\Delta$ and $\mathcal{T}=t/T$, as well as $\bar{h}=\Delta/T$, we have 
 \begin{align}
\tilde{\boldsymbol{P}}\left(\omega\right)&=\epsilon_0\int_{-\infty}^{\infty}\frac{{\rm d}\omega'}{2\pi\bar{h}}\int_{-\infty}^{\infty}{\rm d}\mathcal{T}\,{\rm e}^{\frac{{\rm i}}{\bar{h}}\mathcal{T}(\omega-\omega')}\tilde{\chi}\left(\mathcal{T},\alpha \omega'+(1-\alpha) \omega\right)\tilde{\boldsymbol{E}}\left(\omega'\right)\nonumber\\[10pt]
&=\epsilon_0\hat{\chi}_{\alpha}\tilde{\boldsymbol{E}}\left(\omega\right),\label{eq:operator-form}
\end{align}
$\bar{h}=\Delta/T$ as a dimensionless number that quantifies the timescale of the modulation (`slow' being $\bar{h}\lesssim1$, and `rapid' being $\bar{h}\gg1$).  We now show that $\bar{h}$ is analogous to the constant $\hbar$ in quantum mechanics.  Anticipating this connection, we have written the integral operator in Eq. (\ref{eq:operator-form}) as $\hat{\chi}_{\alpha}$.

In quantum mechanics there is an established method for turning a classical function of phase space $A(p,q)$ into a quantum mechanical operator $\hat{A}$, written as a combination of $\hat{p}$ and $\hat{q}$ operators (for details see Ref.~\cite{Zworski2012}).  As these conjugate operators do not commute, there is an ambiguity in the form of $\hat{A}$ with respect to operator ordering.  For example $A(p,q)=p q^2$ could become $\hat{A}=\hat{p}\hat{q}^2$, $\hat{q}\hat{p}\hat{q}$, $\hat{q}^2\hat{p}$, or a linear combination of these expressions.  The correspondence between the classical function and these different operator orderings is usually parameterized---as we have done in the susceptibility operator (\ref{eq:operator-form})---in terms of a real variable $\alpha$ as follows~\cite{Zworski2012}
\begin{equation}
    \hat{A}_{\alpha}u(q)=\int_{-\infty}^{\infty}{\rm d}q'\int_{-\infty}^{\infty}\frac{{\rm d}p}{2\pi\hbar}{\rm e}^{\frac{\rm i}{\hbar}p(q-q')}\,A(p,\alpha q'+(1-\alpha)q)\,u(q'),\label{eq:operator-quantization}
\end{equation}
i.e. the effect of the operator $\hat{A}_{\alpha}$ on a test function $u$ is given as an integral over phase-space, with the value of $\alpha$ controlling where the function $A$ is evaluated relative to $u$.  For instance, $\alpha=1$ corresponds to both $u$ and $A$ being evaluated at the same position $q'$, whereas $\alpha=0$ leads to $A$ being evaluated at the fixed position $q$.  As an example of this transition between classical phase space and an operator, take the phase space function $A(p,q)=pq$.  Inserting this expression into Eq. (\ref{eq:operator-quantization}) and replacing $p$ with $p\,{\rm e}^{\frac{{\rm i}}{\hbar}p(q-q')}={\rm i}\hbar\,\partial_{q'}{\rm e}^{\frac{{\rm i}}{\hbar}p(q-q')}$ we have the family of operators
\begin{equation}
    A(p,q)=pq\qquad\rightarrow\qquad\hat{A}_{\alpha}u(q)=\left[-{\rm i}\hbar(1-\alpha) q\,\frac{\partial}{\partial q}-{\rm i}\hbar\alpha\,\frac{\partial}{\partial q}q\right]\,u(q')\label{eq:operator-example}.
\end{equation}
This result shows that $\alpha=1$ leads to an operator ordering where momentum operators occur to the left of position, $\alpha=0$ (so-called ``standard quantization'') for position to the left of momentum, and $\alpha=1/2$ for a symmetric ordering (``Weyl quantization''). 

Comparing the above expressions, we see immediately that the time-varying susceptibility defined in Eq. (\ref{eq:operator-form}) is equivalent to the quantization of phase space given in Eq. (\ref{eq:operator-quantization}): \emph{the susceptibility operator $\hat{\chi}_{\alpha}$ is thus the quantization of the 
`classical'' susceptibility $\tilde{\chi}(\mathcal{T},\omega)$ that is a function of the phase space-coordinates $\omega$ and $\mathcal{T}$}.  The quantization scheme for classical mechanics, parameterized by $\alpha$, is equivalent to the choice of microscopic model for the time-varying medium.  Moreover, the ratio of timescales $\bar{h}$ plays the role of Planck's constant, with slow time modulation being analogous to the semiclassical limit of quantum mechanics.


To apply the above quantization scheme and convert the susceptibility function $\tilde{\chi}(\mathcal{T},\omega)$ in Eq. (\ref{eq:operator-form}) into a pseudo-differential operator we write the $\mathcal{T}$ variable in terms of a derivative with respect to $\omega$, replacing $\tilde{\chi}$ with $\hat{\chi}$ within the integral,
\begin{align}
\tilde{\boldsymbol{P}}\left(\omega\right)&=\epsilon_0\int_{-\infty}^{\infty}\frac{{\rm d}\omega'}{2\pi\bar{h}}\int_{-\infty}^{\infty}{\rm d}\mathcal{T}\int_{-\infty}^{\infty}{\rm d}\xi\,\delta(\xi-\omega+\omega')\,\hat{\chi}\left(-{\rm i}\bar{h}\frac{\partial}{\partial\xi},\alpha \omega'+(1-\alpha) \omega\right){\rm e}^{\frac{{\rm i}}{\bar{h}}\mathcal{T}\xi}\tilde{\boldsymbol{E}}\left(\omega'\right)\nonumber\\[10pt]
&=\epsilon_0\left[\int_{-\infty}^{\infty}{\rm d}\omega'\,\hat{\chi}\left({\rm i}\bar{h}\frac{\partial}{\partial\xi},\alpha \omega'+(1-\alpha) \omega\right)\delta(\xi-\omega+\omega')\,\tilde{\boldsymbol{E}}\left(\omega-\xi\right)\right]_{\xi=0}.\label{eq:pseudo-1}
\end{align}
Although expression (\ref{eq:pseudo-1}) looks rather complicated, it makes precise the ordering of frequency derivatives and frequency operators within the susceptibility operator, described in eg. Ref.~\cite{Horsley2023}, analogous to the usual passage from classical to quantum mechanics.  In the following sections we will assume the operator ordering corresponding to $\alpha = 1$, at which point (\ref{eq:pseudo-1}) reduces to
\begin{align}
(\alpha=1):\qquad\widetilde{\boldsymbol{P}} &= \epsilon_0 \int_{-\infty}^{\infty} {\rm d}\omega'\; \widetilde{\boldsymbol{E}}(\omega')\; \hat{\chi}\left(-{\rm i}\bar{h}\frac{\partial}{\partial \omega},\, \omega'\right) \delta(\omega - \omega')\nonumber\\[5pt]
&= \epsilon_0\; \hat{\chi}_{1}\!\left(\hat{\mathcal{T}},\, \hat{\omega}\right) \widetilde{\boldsymbol{E}},\label{eq:chi_L1}
\end{align}
where $\hat{\mathcal{T}}=-{\rm i}\bar{h}\partial_{\omega}$ and we include the subscript `1' to indicate the specific operator order that comes from carrying out the integral in the first line of (\ref{eq:chi_L1}), as introduced in Eq. (\ref{eq:operator-form}).  As the above analysis shows, this $\alpha=1$ operator ordering is a consequence of our chosen form of the time dependence of the susceptibility in (\ref{eq:pseudo-1}).

\begin{figure}[h!]
\begin{center}
\includegraphics[width=0.6\textwidth]{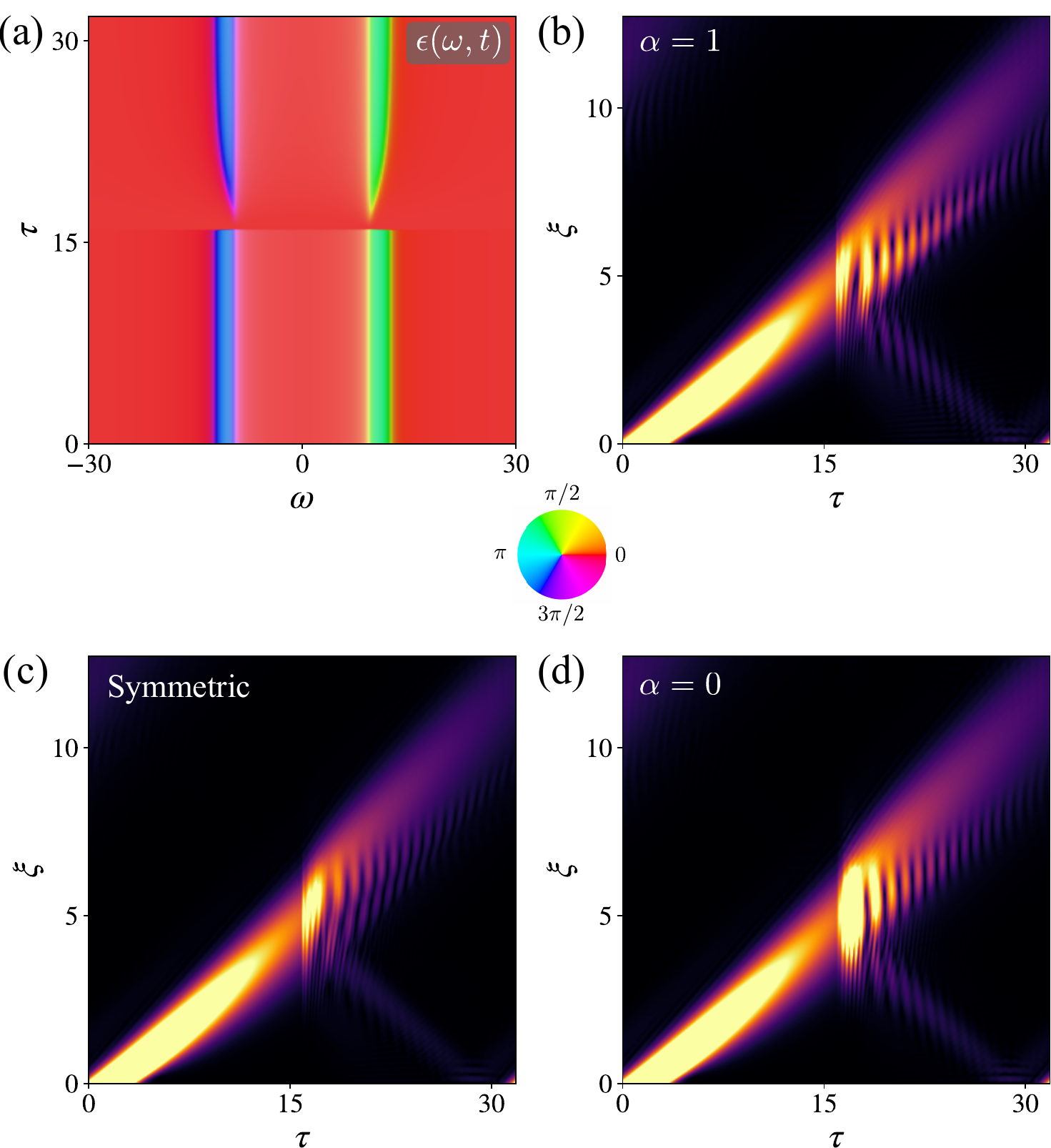}
\end{center}
\caption{\textbf{The effect of operator ordering on temporal scattering:} \textbf{(a)} We consider a Drude-Lorentz model $\varepsilon(\omega,t)=\epsilon_{\rm b}-\omega_{\rm p}^2(t)[\omega^2+{\rm i}\gamma\omega-\omega_0^2]^{-1}$, where the field-matter coupling (the plasma frequency $\omega_p$) is changed in time as $\omega_{\rm p}(t)=\omega_{1}+\frac{\Delta\omega}{2}[1+\tanh((t-t_0)/T)]{\rm e}^{-\lambda (t-t_0)}$, as shown in the complex plot (color is phase, as indicated, saturation is amplitude).  Parameters are $\epsilon_{\rm b}=2.5$, $\gamma=1.5$, $\omega_0=15$, $\omega_{1}=20$, $\Delta\omega=20$, $T=0.02$, and $\lambda=10$. \textbf{(b--d)}  Propagation through the medium computed using the evolution operator $\exp(\frac{{\rm i}\xi}{\bar{h}}\hat{K})$ for a Gaussian pulse $\langle\omega|\psi(0)\rangle=\exp(-a\,(\omega-\omega_c)^2+{\rm i}\omega t_{c})$ at $\xi=0$ ($a=0.2$,$\omega_c=10$, $t_c=1$).  The operator ordering in the permittivity $\hat{\varepsilon}$ depends on the details of how the time modulation is included in the material response (see Appendix). As the figure shows, the choice of operator ordering can lead to very different predictions for the temporal scattering coefficients (see Ref.~\cite{Dixon2026}).\label{fig:scattering}}
\end{figure}

In the regime where the parameter $\bar{h}$ is not small this difference in ordering can produce very different predictions.  An example is shown in Fig.~\ref{fig:scattering}.  In contrast to the situation in quantum theory, we can understand this dependence on the order of the operators in terms of the known model dependence of the scattering coefficients in time varying media~\cite{Solis2021,Galiffi2025,Dixon2026} (see appendix below for details).  For instance, in a Drude-Lorentz model~\cite{Dixon2026,Solis2021}, if the coupling from the field \emph{into} the medium dynamics is time-modulated, the differential operators are evaluated with the frequency derivatives appearing to the \emph{right} of the frequency dependence ($\alpha=0$ in Eq. (\ref{eq:pseudo-1})).  By contrast, a time modulated coupling \emph{out} of the medium into the field~\cite{ptitcyn2023}, results in frequency derivatives appearing on the \emph{left} of all frequency dependence in the susceptibility ($\alpha=1$ in Eq. (\ref{eq:pseudo-1})). In a model based on a Lagrangian that describes the dynamics of both the field and the medium, as suggested in Ref.~\cite{HorsleyBaker2025}, the coupling into and out of the medium appear symmetrically in the susceptibility and -- assuming it is this coupling that is modulated in time -- something similar to Weyl quantization ($\alpha=\frac{1}{2}$) seems more appropriate (although not quite, see appendix).  For simplicity of the numerical treatment we work with \ref{eq:chi_L1} in what follows, remembering that real media may be better described using a different operator ordering.

It is worth contrasting this situation with the well known additional boundary condition (ABC) problem in scattering from \emph{spatially} dispersive media.  As described in e.g.~\cite{Verger2004}, there is some freedom in choosing the boundary conditions between different spatially dispersive media, characterized in terms of different permittivity functions $\epsilon(k)$.  This leads to different predictions for the scattering amplitudes, which has previously been the subject of some debate~\cite{Henneberger1998}.  In light of the above discussion we can see this problem is simply the classical electromagnetic counterpart of (\ref{eq:operator-quantization}) where the permittivity depends on both position and the momentum operators.  Again, different operator orderings arise from different models for the material dynamics, which ultimately leads to different predictions for the behaviour of the wave.  Understanding the permittivity as a quantity that must be ``quantized'' and turned into a pseudodifferential operator, as in the passage from classical to quantum mechanics, makes this situation much clearer.
%
%
\section{Non-Hermitian Schr\"{o}dinger equation for wave propagation}
As discussed above, the use of a frequency domain operator to describe the dielectric susceptibility of a time-varying medium comes with the same choice of operator ordering as we must use in passing from classical to quantum mechanics.  We now show that the evolution of a wave in a time-varying medium is governed by an effective Hamiltonian that depends on this susceptibility operator, which acts to reshape the frequency spectrum (the analogue of the position representation of the wave function in quantum mechanics) of the wave as it propagates.

We start by substituting the operator expression (\ref{eq:chi_L1}) for the polarization into the frequency domain Maxwell's equations.  This results in,
\begin{align}
\bar{h}\boldsymbol{\nabla}_{\xi} \times \widetilde{\boldsymbol{E}} &= {\rm i}\hat{\omega} c\widetilde{\boldsymbol{B}}\nonumber\\
\bar{h}\boldsymbol{\nabla}_{\xi} \times c\widetilde{\boldsymbol{B}} &= -{\rm i}\hat{\omega} \left[\hat{1} + \hat{\chi}_{1}\!\left(\hat{\mathcal{T}},\hat{\omega}\right)\right]\widetilde{\boldsymbol{E}}.
\end{align}
where we have used dimensionless spatial coordinates, e.g.  $\xi = x/cT$, also remembering that we are using the dimensionless time $\mathcal{T}$ and frequency $\omega$ defined above to specify the time dependence of the fields.  Combining these two equations, assuming propagation along the $x$-axis, and taking a single polarization $\tilde{\boldsymbol{E}}=\boldsymbol{e}_{z}\tilde{E}$, we obtain an operator counterpart of the usual one-dimensional Helmholtz equation
\begin{equation}
\left(\bar{h}^2\frac{{\rm d}^2}{{\rm d}\xi^2} + \hat{K}^2\right)\widetilde{E} = 0,\label{eq:op_helm}
\end{equation}
where
\begin{equation}
\hat{K}^2=\hat{\omega}^2\left(1 + \hat{\chi}_{1}\right).\label{eq:K2-chi}
\end{equation}
Assuming that the medium is spatially homogeneous, this operator Helmholtz equation (\ref{eq:op_helm}) can be factorized as follows,
\begin{equation}
\left(\bar{h}\frac{{\rm d}}{{\rm d}\xi} + {\rm i}\hat{K}\right)\!\left(\bar{h}\frac{\rm d}{{\rm d}\xi} - {\rm i}\hat{K}\right)\widetilde{E} = 0,
\end{equation}
where $\hat{K}$ is the operator square root, i.e. $[\hat{K}^{2}]^{1/2}=\hat{K}$. The electric field thus generally obeys a Schr\"odinger-like equation
\begin{equation}
{\rm i}\bar{h}\,\dfrac{{\rm d}}{{\rm d}\xi}|\psi\rangle = \pm\,\underbrace{\hat{K}(\hat{\mathcal{T}},\hat{\omega})}_{\hat{H}}|\psi\rangle,\label{eq:tvm-Schrodinger}
\end{equation}
where the `ket' state is defined such that $\langle\omega|\psi\rangle=\widetilde{E}(x,\omega)$, and the plus/minus sign corresponds to left/right going propagation of the wave (provided the square root of $\hat{K}^2$ is taken appropriately~\cite{Horsley2023}).  Note that these two signs in the Schr\"odinger equation are also present in quantum mechanics when imposing past and future boundary conditions (i.e. forwards and backwards in time propagation) on a quantum system (e.g. Aharonov's two-state vector formalism~\cite{Aharonov1964}).  However, note that---unlike in  conventional quantum mechanics---the Hamiltonian given here is almost always non-Hermitian.  

\begin{figure}[h!]
    \begin{center}
    \includegraphics[width=0.92\textwidth]{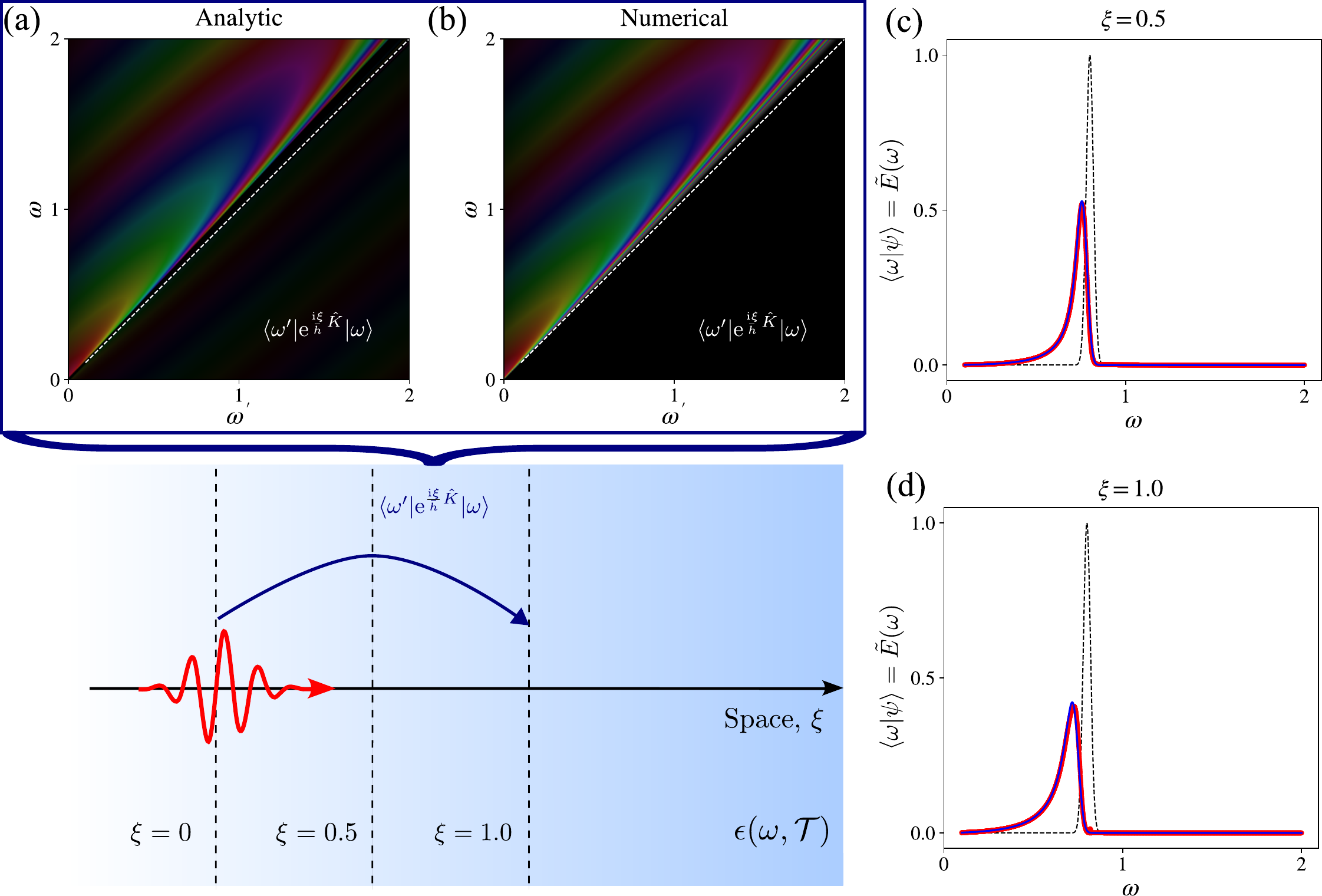}
    \end{center}
    \caption{\textbf{The `propagator' in a time-varying medium:} \textbf{(a--b)} Comparison between the analytic (\ref{eq:example-propagator}) and numerically evaluated propagator $\langle\omega'|\exp(\frac{{\rm i}\xi}{\bar{h}}\hat{K})|\omega\rangle$\label{fig:example-propagation} for $\varepsilon_{\rm b}=3.5$, $\xi=1$ and $\bar{h}=0.5$.  To numerically evaluate the propagator we used a finite (forward) difference approximation to Eq. (\ref{eq:K2-example}) (4500 points), taking the numerical square root and exponential using SciPy's linear algebra library~\cite{scipy} (see Fig. \ref{fig:scattering} for colour scale).  \textbf{(c--d)} Comparison between numerical evolution (blue, solid) and use of analytic propagator (\ref{eq:example-propagator}) (red, dots), showing the redshift and reshaping of the spectrum.  Note that the analytic expression (\ref{eq:example-propagator}) oscillates rapidly around $\omega=\omega'$, making it difficult to evaluate (further work on the asymptotics of the error functions would likely solve this problem).  We used a high frequency filter to remove these high frequency oscillations when numerically evaluating (\ref{eq:example-propagator}).}
\end{figure}

The solution to Eq. (\ref{eq:tvm-Schrodinger}) is formally given by
\begin{equation}
    |\psi(\xi)\rangle={\rm e}^{\mp\frac{{\rm i}\xi}{\bar{h}}\hat{K}}|\psi(0)\rangle,\label{eq:hamiltonian-evolution}
\end{equation}
which, when evaluated in frequency representation, describes how the frequency spectrum $\widetilde{E}(x,\omega)$ of the wave changes as we look at successive depths $\xi$ into the time-varying medium.  See the section below and Fig. \ref{fig:example-propagation} for an example where the `propagator' $\exp(\mp\frac{{\rm i}\xi}{\bar{h}}\hat{K})$ can be evaluated exactly.  When the time modulation is removed, the wave-vector operator becomes diagonal in the frequency basis which, via Eq. (\ref{eq:hamiltonian-evolution}), implies that each frequency component of the wave just receives a phase shift during propagation, as expected in a static medium.  The time modulation appears as frequency derivatives within the operator, the action of which reshapes the frequency spectrum of the wave during propagation.
%
%
%
%
%
%

Given a particular time-varying permittivity, it is very difficult to calculate the analytic expression for the operator square root of $\hat{K}^2$.  As described in~\cite{Horsley2023}, it is often simpler to proceed numerically.  However, there are some situations where we can make analytical progress.  The first is where we can work in terms of the time-varying refractive index $\hat{n}(\hat{\mathcal{T}}, \hat{\omega})$.  We define this such that we can specify the wave-vector operator directly
\begin{equation}
    \hat{K} = \hat{\omega}\, \hat{n}(\hat{\mathcal{T}}, \hat{\omega}).\label{eq:K-op-ind}
\end{equation}
In this case the permittivity operator can be simply found from the refractive index.  Squaring the wave-vector operator (\ref{eq:K-op-ind})
\begin{equation}
\hat{K}^{2} = \hat{\omega}\,\hat{n} \,\hat{\omega}\, \hat{n} = \hat{\omega}^{2} \left(\hat{n}^{2} + \hat{\omega}^{-1}[\hat{n}, \hat{\omega}]\,\hat{n}\right),
\end{equation}
we find that the permittivity operator is given by
\begin{equation}
\hat{\varepsilon} = \hat{n}^{2} + \hat{\omega}^{-1}[\hat{n}, \hat{\omega}]\hat{n}.
\end{equation}
Note that, unlike \textit{in the case of} a static medium, the permittivity operator is not simply the square of the refractive index operator.  The new term, proportional to the commutator of the refractive index and the frequency, represents a time derivative of the refractive index, which is zero for static media.

Another approach is to construct $\hat{K}$ in terms of the eigenvectors and eigenvalues of $\hat{K}^2$.  Given that $\hat{K}^2$ typically depends on a series of terms involving combinations of the frequency $\omega$, and derivatives $\frac{\partial}{\partial\omega}$, we can often find its left and right eigenvectors, $|\phi_\lambda\rangle$ and $|\chi_\lambda\rangle$ obeying
\begin{align}
\hat{K}^2\,|\phi_\lambda\rangle &= \lambda\,|\phi_\lambda\rangle\nonumber\\
\langle\chi_\lambda|\,\hat{K}^2 &= \lambda\,\langle\chi_\lambda|
\end{align}
so that, provided $\langle\chi_\lambda|\phi_\lambda\rangle\neq0$ (i.e. away from exceptional points), the $\hat{K}$ operator appearing in (\ref{eq:tvm-Schrodinger}) can be decomposed as
\begin{equation}
\hat{K} = \sum_{\lambda} \sqrt{\lambda}\,|\phi_\lambda\rangle\langle\chi_\lambda|.
\end{equation}
Here we take the root where $\sqrt{\lambda}$ has a positive imaginary part and a sum is appropriate for cases where the infinite dimensional operator has been truncated to a finite dimensional space.  The sum should be interpreted as an integral otherwise.

%
%
\subsection{Example solution: a dispersive medium with linear time modulation}

Before looking at the semi-classical limit of the above Schr\"{o}dinger theory, we take a simple example where this quantum-like theory can be applied to find the exact propagation of a wave through a time-varying, dispersive medium.  Consider a medium where the permittivity varies linearly in time, $\varepsilon(\tau,\omega) = \varepsilon_b(\omega) +  \bar{h}\tau$, where $\varepsilon_b$ is a function of frequency obeying the Kramers-Kronig relations. The squared wave-vector operator $\hat{K}^2$ is, from Eq.\ (\ref{eq:K2-chi}),
\begin{equation}
\hat{K}^2 = \hat{\omega}^2\!\left(\varepsilon_b(\hat{\omega}) +\hat{\mathcal{T}}\right).\label{eq:K2-example}
\end{equation}
The left and right eigenvectors of this operator are straightforward to find and are given by
\begin{equation}
\langle\omega|\phi_\lambda\rangle = \exp\!\left(-\frac{\rm i}{ \bar{h}}\left[\int_{0}^{\omega}{\rm d}\omega'\varepsilon_b(\omega') +\frac{\lambda}{\omega}\right]\right)\label{eq:right-vector}
\end{equation}
and
\begin{equation}
\langle\chi_\lambda|\omega\rangle = \frac{1}{\omega^2}\exp\!\left(\frac{\rm i}{ \bar{h}}\left[\int_{0}^{\omega}{\rm d}\omega'\varepsilon_b(\omega') + \frac{\lambda}{\omega} \right]\right).\label{eq:left-vector}
\end{equation}
The frequency spectrum (\ref{eq:right-vector}) is an example of a so-called ``eigenpulse'' of the medium~\cite{Horsley2023}, a special wave spectrum that retains its form during propagation in the time-varying medium.  The eigenvectors (\ref{eq:right-vector}--\ref{eq:left-vector}) also form a complete set, with the completeness relation
\begin{equation}
\int_{-\infty}^{\infty} \frac{{\rm d}\lambda}{2\pi \bar{h}}\,\langle\omega|\phi_\lambda\rangle\langle\chi_\lambda|\omega'\rangle = \delta(\omega - \omega').
\end{equation}
The `propagator', giving the frequency spectrum of the field after propagating through a spatial distance $\xi$, can be constructed from the above completeness relation
\begin{align}
\langle\omega'|{\rm e}^{\frac{{\rm i}\xi}{\hbar}\hat{K}}|\omega\rangle &= \int_{-\infty}^{\infty}\frac{{\rm d}\lambda}{2\pi \bar{h}}\,\langle\omega'|\phi_\lambda\rangle\langle\chi_\lambda|\omega\rangle\,{\rm e}^{\frac{{\rm i}\xi}{\bar{h}}\sqrt{\lambda}}\nonumber \\[5pt]
&=\frac{1}{\pi\bar{h}\omega^2}{\rm e}^{\frac{\rm i}{\bar{h}}\int^{\omega}_{\omega'}{\rm d}\omega_1\,\epsilon_{\rm b}(\omega_1)}\lim_{\eta\to0}\left[\int_0^{\infty}k\,{\rm d}k\,{\rm e}^{\frac{\rm i}{\bar{h}}k^2\left(\frac{1}{\omega}-\frac{1}{\omega'}+{\rm i}\eta\right)}{\rm e}^{\frac{{\rm i}\xi}{\bar{h}}k}-\int_{0}^{{\rm i}\infty}k\,{\rm d}k\,{\rm e}^{\frac{\rm i}{\bar{h}}k^2\left(\frac{1}{\omega}-\frac{1}{\omega'}-{\rm i}\eta\right)}{\rm e}^{\frac{{\rm i}\xi}{\bar{h}}k}\right]\nonumber\\[5pt]
&=\delta(\omega-\omega')\nonumber\\[5pt]
&+\frac{\xi{\rm e}^{\frac{\rm i}{\bar{h}}\int_{\omega'}^{\omega}\epsilon_{\rm b}(\omega_1){\rm d}\omega_1}}{4\omega^2\sqrt{\pi\bar{h}}}\lim_{\eta\to0}\left[\left(\frac{1}{{\rm i}X}\right)^{3/2}{\rm i}\,{\rm erfc}\left(\frac{-{\rm i}\xi}{2\sqrt{{\rm i}\bar{h}X}}\right){\rm e}^{\frac{{\rm i}\xi^2}{4\bar{h}X}}-\left(\frac{1}{-{\rm i}X^{\ast}}\right)^{3/2}{\rm erfc}\left(\frac{\xi}{2\sqrt{-{\rm i}\bar{h}X^{\ast}}}\right){\rm e}^{\frac{{\rm i}\xi^2}{4\bar{h}X^{\ast}}}\right]\label{eq:example-propagator},
\end{align}
where $X=(1/\omega' - 1/\omega - {\rm i}\eta)$.  The above result (\ref{eq:example-propagator}) has been written in terms of a sum of two complementary error functions, the definitions for which can be found in e.g. Ref.~\cite{NIST:DLMF}.  This exact solution to the Schr\"odinger equation (\ref{eq:tvm-Schrodinger}) is compared with a numerical solution in Fig.~\ref{fig:example-propagation}.  We observe that the complementary error functions--which, as a function of a real variable are Gaussian filtered step functions---serve to spread the frequency spectrum of the wave away from the diagonal ($\omega=\omega'$) in only one direction, in this case to lower frequencies due to the temporally increasing permittivity in Eq. (\ref{eq:K2-example}).
%
%
\begin{figure}[h!]
\begin{center}
\includegraphics[width=0.75\textwidth]{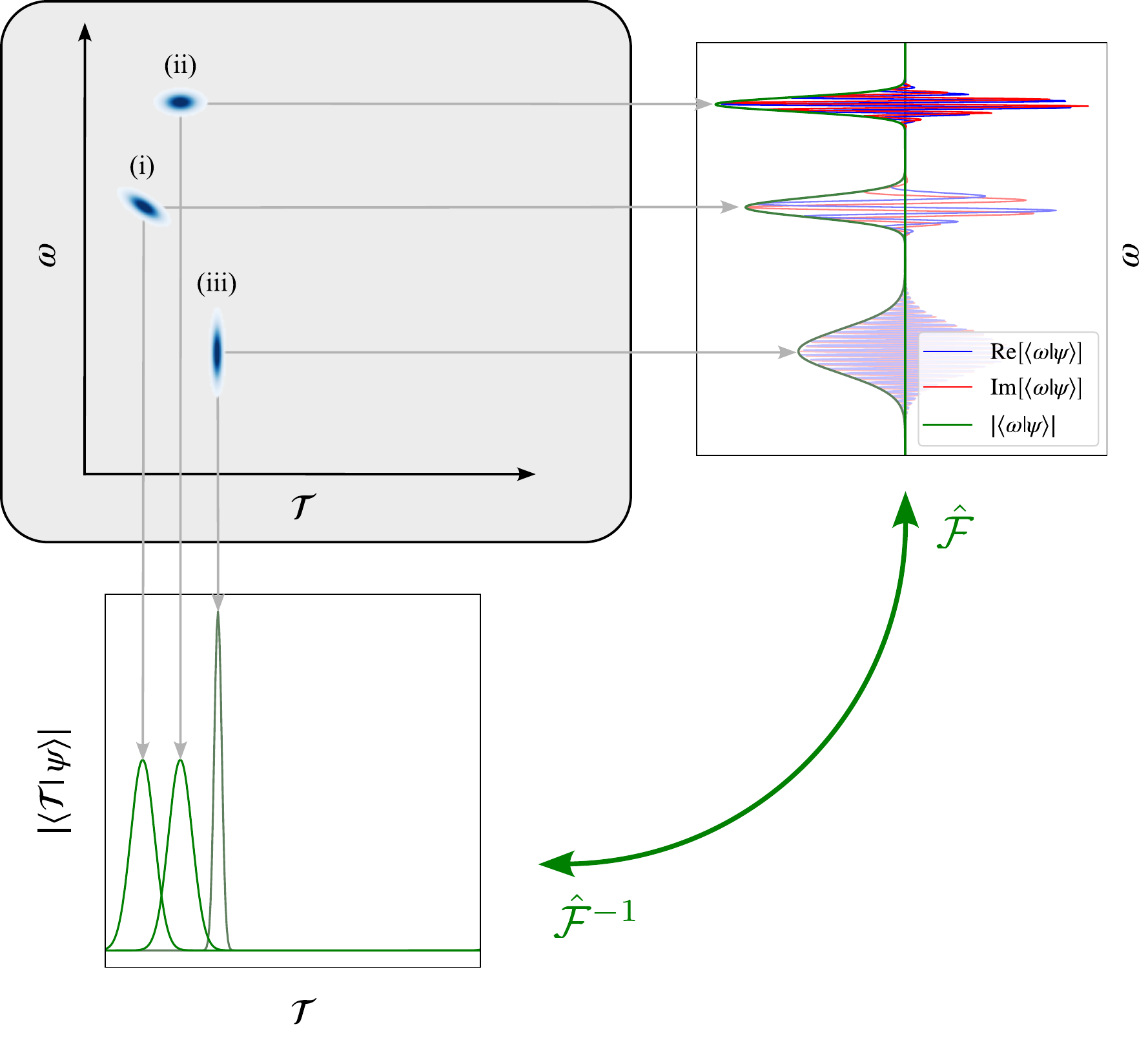}
\end{center}
\caption{\textbf{Using the Wigner function to visualize wave motion:}  As defined in Eq. (\ref{eq:Wigner-def}), the Wigner function $W(\omega,\mathcal{T})$ can be understood as a quasi-power spectrum in \emph{both} frequency and time, giving us information of both the frequency content and the arrival time of any given signal.  The upper left panel shows a plot of the Wigner function for three different Gaussian time domain signals: (i) a pulse chirped in frequency/time, which is evident as a rotated distribution in phase space; (ii) a pulse squeezed in frequency; and (iii) a pulse squeezed in time.  The right and lower panels show the frequency and time domain representations of these spectra, respectively, with the chirp of distribution (i) evident in the real/imaginary parts of the frequency domain dependence.\label{fig:wigner-function}}
\end{figure}
\section{The Wigner function and the ``semi-classical'' limit:\label{sec:wigner}}
Given the parallel with quantum theory, we can apply the apparatus introduced in Sec.~\ref{sec:ordering} used to connect classical phase-space and quantum mechanics to take the ``semiclassical'' limit where $\bar{h} \ll 1$, which in our case indicates a large separation of timescales between the wave period and the modulation time. In this regime it becomes appropriate to use the notion of a trajectory in phase space, where the central arrival time and frequency smoothly evolve along a curve, with increasing $\xi$.  An example of such a trajectory is sketched in Fig. \ref{fig:phase-space}b.

To find the equations of motion governing this trajectory, we follow Ref.~\cite{graefe2011,graefe2012,graefe2015} and work in terms of the Wigner function $W$, a quantity often used in quantum mechanics to visualize the state $|\psi\rangle$ as a distribution in phase space rather than as a vector in Hilbert space. This quantity has been used previously in classical optics to describe momentum/position distributions of light fields, and to characterize ultra-short pulses~\cite{paye1992,alonso2004}, but here we use it to describe frequency and time shifts of radiation passing through a time-varying medium.  We introduce this function in terms of a pair of frequency co-ordinates, $\omega_{1}=\omega + \frac{w}{2}$ and $\omega_{2}=\omega - \frac{w}{2}$, with $\omega$ the average of the two co-ordinates and $w$ their difference. We define the Wigner function $W$ as
\begin{equation}
W(\omega, \mathcal{T}) = \int_{-\infty}^{\infty} \frac{{\rm d}w}{2\pi\bar{h}} \left\langle \omega - \frac{w}{2} \Big| \psi \right\rangle \left\langle \psi\Big| \omega + \frac{w}{2} \right\rangle {\rm e}^{\frac{{\rm i}w\mathcal{T}}{\bar{h}}}.\label{eq:Wigner-def}
\end{equation}
Performing an inverse Fourier transform, we can invert this relation to find the density matrix $\hat{\rho}=|\psi\rangle\langle\psi|$ from the Wigner function
\begin{equation}
    \hat{\rho}\,u(\omega)=\int_{-\infty}^{\infty}{\rm d}\omega'\int_{-\infty}^{\infty}{\rm d}\mathcal{T}\,{\rm e}^{{\rm i}(\omega-\omega')\mathcal{T}}W\left(\frac{\omega+\omega'}{2},\mathcal{T}\right)u(\omega'),
\end{equation}
which is simply the formula (\ref{eq:operator-quantization}) used to move between operators and phase-space functions introduced in Sec.~\ref{sec:ordering}.  For the particular case of a Gaussian frequency spectrum $\langle\omega|\psi\rangle=({\rm Re}[a]/\pi)^{1/4}\exp\left(-\frac{a}{2}(\omega-\omega_0)^2+\frac{{\rm i}\mathcal{T}_0}{\bar{h}}(\omega-\omega_0)\right)$, the Wigner function (\ref{eq:Wigner-def}) reduces to
\begin{equation}
    \text{Gaussian state:}\qquad W(\omega,\mathcal{T})=\frac{1}{\pi\bar{h}}\exp\left[-\frac{1}{\bar{h}}(Z-Z_0)\cdot\Sigma^{-1}\cdot (Z-Z_0)\right]\label{eq:W_gaussian},
\end{equation}
where
\begin{equation}
    \Sigma^{-1}=\frac{1}{{\rm Re}[a]}\left(\begin{matrix}\bar{h}^{-1}&\frac{1}{2}{\rm Im}[a]\\\frac{1}{2}{\rm Im}[a]&\bar{h}\left({\rm Re}[a]^2+\frac{1}{4}{\rm Im}[a]^2\right)\end{matrix}\right),\label{eq:covariance-inv}
\end{equation}
and we have introduced the vector $Z=(\mathcal{T},\omega)^{\rm T}$.  Fig.~\ref{fig:wigner-function} shows examples of the Wigner function (\ref{eq:W_gaussian}), where we can see that the duration, bandwidth, and chirp of a pulse depend on the real and imaginary parts of $a$, and can be simply visualized in frequency-time phase space.  Note that the matrix $\Sigma$ is $2/\bar{h}$ times the covariance matrix of the Gaussian distribution $|\langle\omega|\psi\rangle|^2$.

We now use the non-Hermitian Schr\"{o}dinger equation (\ref{eq:tvm-Schrodinger}) derived above to compute the evolution of this phase space distribution $W(\omega,\mathcal{T})$ with increasing propagation depth $\xi$.  The wave-vector operator is generally non-Hermitian, $\hat{K} = \hat{K}_1 \pm {\rm i}\,\hat{K}_2$, with both $\hat{K}_1$ and $\hat{K}_2$ Hermitian.  The time evolution of the density operator then depends on both the commutator $[\hat{K}_{1},\hat{\rho}]$ and anti-commutator $\{\hat{K}_{2},\hat{\rho}\}$
\begin{equation}
    {\rm i}\bar{h}\frac{{\rm d}\hat{\rho}}{{\rm d}\xi}=-\left[\hat{K}_1,\hat{\rho}\right]-{\rm i}\left\{\hat{K}_{2},\hat{\rho}\right\}
\end{equation}
where we have assumed rightwards propagation in Eq. (\ref{eq:tvm-Schrodinger}).  Transforming from operator to phase-space distribution using Eq. (\ref{eq:Wigner-def}), we find that the Wigner function evolves as
\begin{equation}
{\rm i}\bar{h} \frac{{\rm d}}{{\rm d}\xi} W(\omega, \mathcal{T}) = -\int_{-\infty}^{\infty} \frac{{\rm d}w}{2\pi\bar{h}} \left\langle \omega - \frac{w}{2} \Big| \left([\hat{K}_1, \hat{\rho}] + i\{\hat{K}_2, \hat{\rho}\}\right) \Big| \omega + \frac{w}{2} \right\rangle {\rm e}^{\frac{{\rm i}w\mathcal{T}}{\bar{h}}}.\label{eq:wigner-function-propagation}
\end{equation}
To evaluate the right-hand side of (\ref{eq:wigner-function-propagation}) we rewrite the integral of the operator products in terms of the Wigner function.  To do this we use the following Moyal product for arbitrary operators $\hat{A}$ and $\hat{B}$~\cite{hillery1984}
\begin{align}
\int_{-\infty}^{\infty} {\rm d}w \left\langle \omega - \frac{w}{2} \Big| \hat{A}\hat{B} \Big| \omega + \frac{w}{2} \right\rangle {\rm e}^{\frac{{\rm i}\,w\,\mathcal{T}}{\bar{h}}} &= A(\omega, \mathcal{T}) \exp\!\left(\frac{{\rm i}\bar{h}}{2}\left(\frac{\overleftarrow{\partial}}{\partial \omega}\frac{\overrightarrow{\partial}}{\partial \mathcal{T}} - \frac{\overleftarrow{\partial}}{\partial \mathcal{T}}\frac{\overrightarrow{\partial}}{\partial \omega}\right)\right) B(\omega, \mathcal{T})\nonumber\\[10pt]
&=A(Z)\exp\left(\frac{{\rm i}\bar{h}}{2}\overleftarrow{\nabla}\cdot\Omega\cdot\overrightarrow{\nabla}\right)B(Z)\label{eq:moyal-product}.
\end{align}
In the second line we have used a compressed notation where $\nabla=(\partial_{\mathcal{T}},\partial_{\omega})^{\rm T}$ and $\Omega=\left(\begin{matrix}0&-1\\1&0\end{matrix}\right)$, and the distribution functions are related to the operators by
\begin{equation}
    A(\omega,\mathcal{T})=\int_{-\infty}^{\infty}{\rm d}w\left\langle\omega-\frac{w}{2}\Big|\hat{A}\Big|\omega+\frac{w}{2}\right\rangle\,{\rm e}^{\frac{{\rm i}\,w\,\mathcal{T}}{\bar{h}}}.
\end{equation}
Applying the result (\ref{eq:moyal-product}) to Eq. (\ref{eq:wigner-function-propagation}) and expanding the right hand side up to order $\bar{h}$, the distribution of our field in phase space (the Wigner function) obeys
\begin{equation}
    \frac{{\rm d}W}{{\rm d}\xi}=-\left[K_1(\overleftarrow{\nabla}\cdot\Omega\cdot\overrightarrow{\nabla})+\frac{2}{\bar{h}}K_{2}-\frac{\bar{h}}{4}K_2(\overleftarrow{\nabla}\cdot\Omega\cdot\overrightarrow{\nabla})^2\right]W.
\end{equation}
This differential equation governs the evolution of the phase-space quasi-probability distribution $W$ in leading order of $\bar{h}$.  Inserting the Wigner function for the Gaussian state (\ref{eq:W_gaussian}), assuming a time dependent $Z_0$ and $\Sigma$, and equating like powers of $(Z-Z_0)$ yields the coupled equations of motion
\begin{align}
    \frac{{\rm d}Z_{0}}{{\rm d}\xi}&=K_{1}'\cdot\Omega-K_2'\cdot\Sigma\nonumber\\[10pt]
    \frac{{\rm d}\Sigma}{{\rm d}\xi}&=\Sigma\cdot K_{1}''\cdot\Omega-\Omega\cdot K_{1}''\cdot\Sigma-\Sigma\cdot K_{2}''\cdot\Sigma-\Omega\cdot K_{2}''\cdot\Omega,\label{eq:hamilton}
\end{align}
where the wave-vectors $K_{1}$ and $K_2$ have been expanded up to second order around the centre of the wave-packet at $Z_0$, e.g. $K_{1}=K_1^{(0)}+K_{1}'\cdot(Z-Z_0)+\frac{1}{2}(Z-Z_0)\cdot K_{1}''\cdot(Z-Z_0)$.  Analogues of the above equations are given for a quantum system in Ref.~\cite{graefe2011,graefe2015}, and in our case   tell us that the centre of a wave-packet's frequency-time distribution moves according to a generalized version of Hamilton's equations
\begin{align}
    \frac{{\rm d}\mathcal{T}_0}{{\rm d}\xi}&=\frac{\partial K_{1}}{\partial\omega_{0}}-\frac{\partial K_2}{\partial\mathcal{T}_{0}}\Sigma_{11}-\frac{\partial K_2}{\partial\omega_{0}}\Sigma_{21}\nonumber\\[10pt]
    \frac{{\rm d}\omega_0}{{\rm d}\xi}&=-\frac{\partial K_{1}}{\partial\mathcal{T}_{0}}-\frac{\partial K_2}{\partial\mathcal{T}_{0}}\Sigma_{12}-\frac{\partial K_2}{\partial\mathcal{\omega}_0}\Sigma_{22}.\label{eq:classical-equations}
\end{align}
Note that a similar generalization of Hamilton's equations can alternatively be been derived using a Gaussian ansatz for the position representation of the wavefunction without reference to the Wigner function~\cite{graefe2012}.  Without dissipation ($K_2=0$) these equations are simply an expression of the spatial translational invariance of the time-varying medium, where $K$ is conserved, i.e. $\frac{{\rm d}K}{{\rm d}\xi}=\frac{{\rm d}\mathcal{T}_{0}}{{\rm d}\xi}\frac{\partial K_1}{\partial\mathcal{T}_0}+\frac{{\rm d}\omega_{0}}{{\rm d}\xi}\frac{\partial K_1}{\partial\omega_0}=0$.  This is the direct analogue of the situation in classical mechanics where Hamilton's equations tell us that, in a non-dissipative time translational invariant system, \emph{energy} is conserved.  However, when dissipation is non-negligible ($K_2\neq0$) the dynamics becomes more complex and depends on the phase-space distribution of the field through the matrix $\Sigma$. 

Something to note in passing is that the expansion of the wave-vector in Eq. (\ref{eq:hamilton}) requires it to be analytic around the expansion point.  This assumption may fail in the vicinity of a point where the permittivity vanishes, where there is a square root branch point in the wave-vector.  In practice, dissipation usually moves such points significantly away from the real frequency axis.  However, this non-analyticity may be relevant in systems with low dissipation.  Assuming the expansion of the wave-vector \emph{can} be carried out, we now explore the use of the equations of motion (\ref{eq:classical-equations}) to describe the evolution of a wave-packet in a dispersive time varying medium.  

\begin{figure}[h!]
\includegraphics[width=0.9\textwidth]{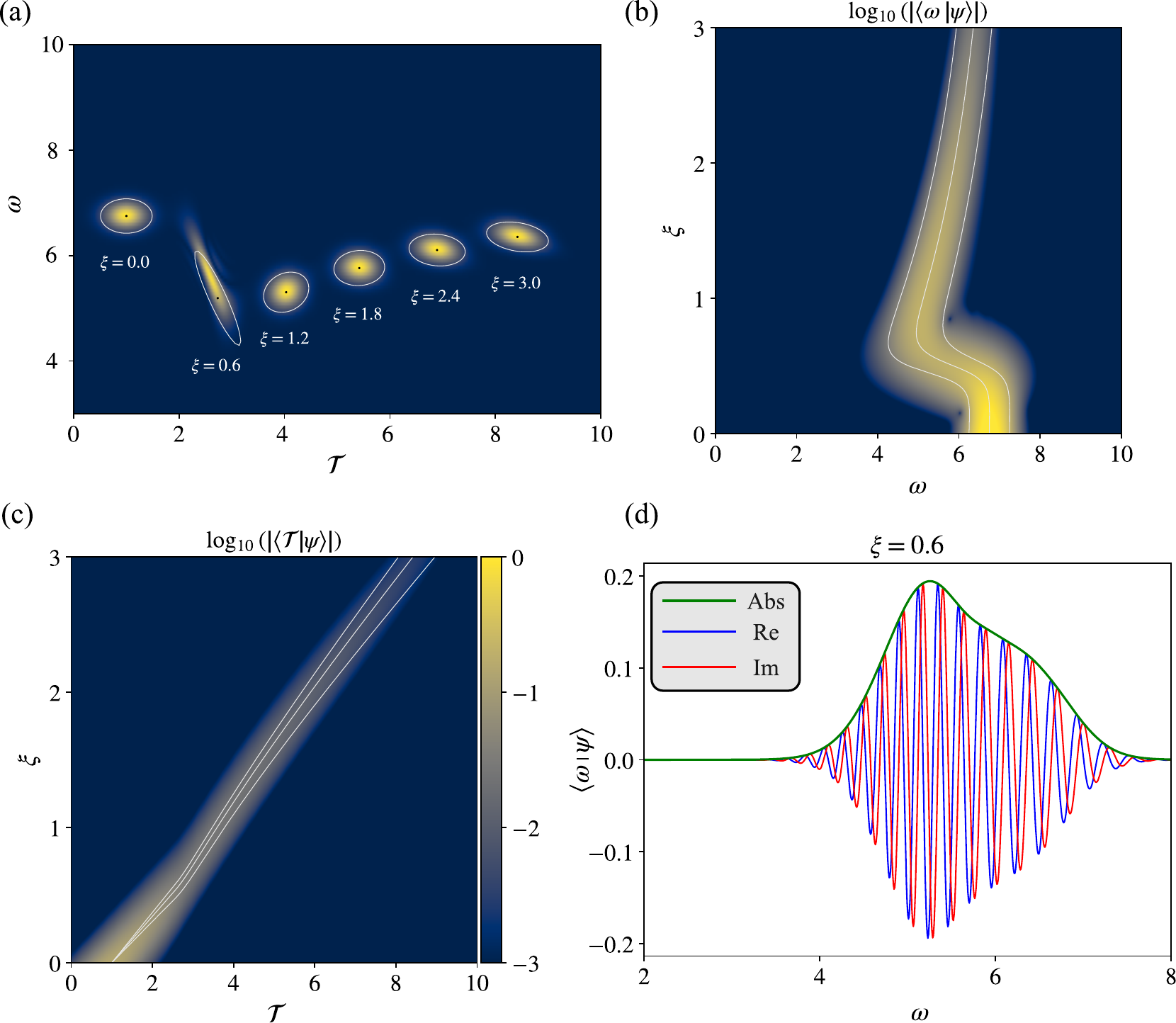}
\caption{\textbf{Semiclassical evolution of a wave-packet in a dispersive time-varying medium:}   We take a Drude model with a time-varying plasma frequency (see appendix~\ref{app:drude-details} for permittivity and parameters).  The colourmaps show the absolute value of (b) the frequency and (c) time dependence of the field as we move deeper into the medium, the time modulation causing a significant red shift at around $\xi=0.5$.  Along with this there is (a) the associated Wigner function, computed using Eq. (\ref{eq:Wigner-def}).  Upon each of the plots (a--c) we overlay the frequency-time trajectory and equi-intensity contour $(Z-Z_0)\cdot\Sigma^{-1}\cdot(Z-Z_0)={\rm const.}$, found from the extended Hamilton equations (\ref{eq:hamilton}).  One notable feature of the Wigner function shown in (a) is the significant chirp of the field predicted around $\xi=0.6$.  In panel (d) we plot the complex frequency spectrum of the field at this depth, showing a decreasing rate of oscillation with increasing $\omega$, consistent with the chrip shown in panel (a).~\label{fig:semiclassical}}
\end{figure}

In Fig. \ref{fig:semiclassical} we give an example showing that these equations are a good approximation to the evolution of the electromagnetic field in a time-varying Drude metal, where the wave-vector $K=K_{1}+{\rm i}K_{2}$ is defined as the following distribution in phase space
\begin{equation}
K(\mathcal{T},\omega)=\omega\sqrt{\epsilon_{\rm b}-\frac{\omega_{\rm p}^2(\mathcal{T})}{\omega(\omega+{\rm i}\gamma)}}=\omega\sqrt{\epsilon_{\rm b}-\frac{\omega_{\rm p}^2(\mathcal{T})}{(\omega^2+\gamma^2)}+\frac{{\rm i}\gamma\omega_{\rm p}^2(\mathcal{T})}{\omega(\omega^2+\gamma^2)}}.
\end{equation}
More details are given in Appendix~\ref{app:drude-details}.  

%
%
%
%
%

The equations of motion (\ref{eq:classical-equations}) take a particularly simple form when the $\Sigma$ matrix can be approximated to reduce to the identity matrix
\begin{align}
    \frac{{\rm d}\mathcal{T}_{0}}{{\rm d}\xi}&=\frac{\partial K_{1}}{\partial\omega_0}-\frac{\partial K_2}{\partial\mathcal{T}_{0}}\nonumber\\
    \frac{{\rm d}\omega_0}{{\rm d}\xi}&=-\frac{\partial K_1}{\partial\mathcal{T}_0}-\frac{\partial K_2}{\partial \omega_0}.
\end{align}
These equations correspond to the intuitive picture, where either an increase in the refractive index with time, or increasing loss with frequency, lead to a red shift in the spectrum with increasing propagation distance, ${\rm d}\omega_0/{\rm d}\xi<0$.  Similarly, the rate of change of $K_1$ with frequency tells us the inverse group velocity, which tells us how the pulse arrival time varies with depth, and an increasing loss with time $\partial K_2/\partial\mathcal{T}_0>0$ shifts the arrival time of a pulse to earlier times.  Looking at Eq. (\ref{eq:covariance-inv}), we see this particular case corresponds to a Gaussian state with $a=\bar{h}^{-1}$, i.e.
\begin{equation}
    \langle\omega|\psi\rangle=\left(\frac{1}{\bar{h}\pi}\right)^{1/4}\exp\left(-\frac{1}{2\bar{h}}(\omega-\omega_0)^2+\frac{{\rm i}\mathcal{T}_0}{\bar{h}}(\omega-\omega_0)\right)\label{eq:coherent-gaussian}
\end{equation}
 This family of Gaussian states (\ref{eq:coherent-gaussian}) are frequency-time analogues of coherent states used in the study of the quantum simple harmonic oscillator~\cite{Schrodinger1926}, quantum electromagnetism~\cite{Glauber1963}, and condensed matter physics~\cite{klauder1985}.  Note that Gabor previously introduced these frequency-time coherent states shortly after the Second World War, as ``elementary signals'' in communication theory~\cite{gabor1946}.  These are minimum uncertainty wave-packets that saturate the uncertainty relation
 \begin{equation}
\Delta\omega\; \Delta\mathcal{T} \geq \frac{1}{2}\left|\left[\omega,-{\rm i}\bar{h}\frac{\partial}{\partial\omega}\right]\right|=\frac{\bar{h}}{2}.\label{eq:uncertainty}
\end{equation}
 and are eigenfunctions of the annihilation operator $\hat{a}=(2\bar{h})^{-1/2}(\omega+\bar{h}\partial_{\omega})$
 \begin{equation}
 \hat{a}|\psi\rangle=\frac{1}{\sqrt{2\bar{h}}}\left(\omega_0+{\rm i}\mathcal{T}_{0}\right)|\psi\rangle
\end{equation}
where the eigenvalues are complex representation of the centre of the wave-packet in phase space.  It may be advantageous to characterize time-varying media in the coherent state basis rather than the frequency basis, as is commonly done.
\section{Summary and conclusions}
As emphasized in previous work, it is difficult to understand the propagation of a wave-packet in a medium that is both dispersive and has properties that vary in time.  Yet, this is the situation encountered in recent experiments such as those reported in~\cite{harwood2025,Hooper2025}.  Here we have tried to make progress on this problem, showing that wave propagation can be understood in terms of phase space, where the conjugate variables are the central frequency and arrival time of the wave-packet, which both evolve with propagation distance into the medium.  The analogue of the Hamiltonian (the energy for a system with time translational symmetry) is the wave-vector, which is conserved in a lossless medium with spatial translational symmetry.

To quantify the time modulation, we introduced a dimensionless parameter $\bar{h}$ that is the ratio of the modulation and wave oscillation timescales.  Just as in quantum mechanics, when $\bar{h}$ is small, the frequency and arrival time can be approximated as being simultaneously known and a generalization of Hamilton's equations of motion (\ref{eq:hamilton}) can be used to predict the evolution of a wave-packet as it propagates through the medium.  Meanwhile, when $\bar{h}$ is of order unity, we must use an operator description, where the frequency and arrival time become non-commuting operators, obeying $[\hat{\omega},\hat{\mathcal{T}}]={\rm i}\bar{h}$.  In this regime, we found the spectrum of the wave evolves through the medium according to a non-Hermitian Schr\"odinger equation (\ref{eq:tvm-Schrodinger}).We provided an example of an exact solution for propagation in a medium that is dispersive and varies linearly in time. 

We noted, again just as in quantum mechanics, there is an ambiguity in the ordering of the phase-space variables, $\omega$ and $\mathcal{T}$ in the evolution equation (\ref{eq:tvm-Schrodinger}), when these become operators.  Here we have shown that this ambiguity has a physical interpretation, arising from different microscopic models for the material dynamics.  It was recently shown (e.g.~\cite{HorsleyBaker2025,Dixon2026}) that these model choices can have a dramatic effect on the predictions of the theory when the time modulation is rapid, just as operator ordering has a dramatic effect in quantum theory, when $\hbar$ cannot be taken as small.

Finally, through introducing the frequency-time analogue of the Wigner function (\ref{eq:Wigner-def}) we developed an evolution equation (\ref{eq:wigner-function-propagation}) that can be expanded in powers of $\bar{h}$ to understand wave propagation in terms of a distribution in phase space.  Assuming a Gaussian state with a varying centre and covariance, and considering the leading-order terms in $\bar{h}$, this leads to a generalization of Hamilton's equations (\ref{eq:hamilton}) that well describe the evolution of the wave spectrum in a time-modulated Drude metal (see Fig.~\ref{fig:wigner-function}).


These findings provide a straightforward theoretical approach for analysing wave propagation in dispersive, time-varying media. Additionally,  they also connect the fields of non-Hermitian quantum mechanics and time varying media.  Unlike in quantum mechanics, where a non-Hermitian Hamiltonian is only ever an approximation, here the effective Hamiltonian is nearly always non-Hermitian.  In addition, the phase-space trajectories are usually open in time-varying media, as causality implies the arrival time of a wave-packet increases as we move through the medium, a requirement that is not present in classical or conventional quantum mechanics.  It may be interesting to study this further.

\acknowledgements
S.A.R.H. and E. G. acknowledge financial support from the EPSRC through the program grant ``Next generation metamaterials: exploiting four dimensions'' (Grant No. EP/Y015673/1).  S.A.R.H. acknowledges additional support through the EPSRC grant ``Photon management in dynamic complex scattering media'' (EP/Z535928/1) as well as useful conversations with M. Zworski, J. E. Galkowski, A. Zayats and C. M. Hooper.  In particular S.A.R.H thanks Steve Barnett for suggesting the use of the $Q$ function to analyse propagation, and Zeki Hayran for insightful comments on the first draft.

\bibliography{refs}
\appendix
\section{Different temporal orderings in the Drude-Lorentz model}

In the Drude-Lorentz model, the polarization field behaves as a driven, damped harmonic oscillator. In an undriven, static medium the dynamics of the polarization field are governed by
\begin{equation}
\frac{\partial^{2} \boldsymbol{P}}{\partial t^{2}} + \gamma \frac{\partial \boldsymbol{P}}{\partial t} + \omega_{0}^{2} \boldsymbol{P} = \epsilon_{0} \omega_{p}^{2} \boldsymbol{E},\label{eq:P-dynamics}
\end{equation}
where $\omega_{\rm p}$ is the plasma frequency.  In this case the forcing of the polarization field is equal to $\varepsilon_0\omega_{\rm p}^2\boldsymbol{E}$, and the coupling out (back into the field) is through the current, $\boldsymbol{j} = \frac{\partial \boldsymbol{P}}{\partial t}$ which appears in Maxwell's equations as follows
\begin{equation}
\boldsymbol{\nabla} \times \boldsymbol{H} = \frac{\partial \boldsymbol{P}}{\partial t} + \epsilon_{0} \frac{\partial \boldsymbol{E}}{\partial t}.
\end{equation}
Suppose we now make the medium time-varying through an external modulation of the plasma frequency.  If we simply make the substitution $\omega_p \rightarrow \omega_p(t)$ in Eq. (\ref{eq:P-dynamics}) as in Ref.~\cite{Solis2021} then the solution to the differential equation is,
\begin{equation}
\tilde{\boldsymbol{P}} = \frac{1}{\omega_{0}^{2} - \omega^{2} - {\rm i}\gamma\omega}\, \epsilon_{0}\, \hat{\omega}_{p}^{2}(-{\rm i}h\partial_{\omega})\, \tilde{\boldsymbol{E}} = \hat{\chi}_{R}\tilde{\boldsymbol{E}} \qquad\qquad (\alpha=0).\label{eq:chi_L}
\end{equation}
This assumption -- that the coupling into the polarization field is time-modulated, but the coupling back out \emph{into} the field is not -- is equivalent to making the choice $\alpha=0$ in Eq. (\ref{eq:pseudo-1}). In this case the frequency derivatives in the susceptibility operator sit to the \emph{right} of any frequency dependence.

However, we could also make a different choice. For example, defining $\boldsymbol{P} = \epsilon_0\,\omega_{\rm p}^2(t)\boldsymbol{X}$ with $\boldsymbol{X}$ satisfying
\begin{equation}
\frac{\partial^{2} \boldsymbol{X}}{\partial t^{2}} + \gamma \frac{\partial \boldsymbol{X}}{\partial t} + \omega_{0}^{2} \boldsymbol{X} = \boldsymbol{E},\label{eq:X-dynamics}
\end{equation}
we find the Fourier spectrum of the polarization field to be,
\begin{equation}
\tilde{\boldsymbol{P}} = \epsilon_{0}\, \hat{\omega}_{\rm p}^{2}(-{\rm i}h\partial_{\omega})\!\left(\frac{1}{\omega_{0}^{2}-\omega^{2}-{\rm i}\gamma\omega}\,\tilde{\boldsymbol{E}}\right) \qquad\qquad (\alpha=1),\label{eq:chi_R}
\end{equation}
which is equivalent to choosing $\alpha=1$ in Eq. (\ref{eq:pseudo-1}) (note that (\ref{eq:chi_L}) and (\ref{eq:chi_R}) become identical and both are solutions to (\ref{eq:P-dynamics}) when the time-modulation is removed). In this case the time-modulation is through the coupling \emph{out} of rather than \emph{into} the polarization field.

Neither (\ref{eq:chi_L}) nor (\ref{eq:chi_R}) represents the true situation, where time-modulating the coupling between the field and the medium modifies the coupling of both the field into the material dynamics \emph{and} this dynamics back into the electromagnetic field.  We can implement such a symmetrical time-modulation as follows. We define $\boldsymbol{P} = \epsilon_0\omega_{\rm p}(t)\boldsymbol{X}$ and modify (\ref{eq:X-dynamics}) to
\begin{equation}
\frac{\partial^{2} \boldsymbol{X}}{\partial t^{2}} + \gamma \frac{\partial \boldsymbol{X}}{\partial t} + \omega_{0}^{2} \boldsymbol{X} = \omega_{\rm p}(t)\,\boldsymbol{E}.\label{eq:X2-dynamics}
\end{equation}
Again we can solve (\ref{eq:X2-dynamics}), and the Fourier spectrum of the polarization field is
\begin{equation}
\tilde{\boldsymbol{P}} = \epsilon_{0}\, \hat{\omega}_{\rm p}(-{\rm i}h\partial_{\omega})\, \frac{1}{\omega_{0}^{2}-\omega^{2}-{\rm i}\gamma\omega}\, \hat{\omega}_{p}(-{\rm i}h\partial_{\omega})\tilde{\boldsymbol{E}}.\label{eq:sym-sol}
\end{equation}
We have now imposed a symmetrical ordering of the frequency derivative operators (see Fig. \ref{fig:scattering}), which represents the symmetrical time-modulation of the coupling into and out of the material dynamics.

Note that, despite the symmetrical ordering of the operators in (\ref{eq:sym-sol}), this is not the $\alpha=\frac{1}{2}$ case (Weyl quantization) described in the main text. We can see this in the example case where $\omega_p(t) = \omega_{p0} + \frac{at}{T}$. In this case (\ref{eq:sym-sol}) reduces to
\begin{equation}
\tilde{\boldsymbol{P}} = \epsilon_{0}\left(\omega_{p0} - {\rm i}ha\frac{\partial}{\partial\omega}\right)\frac{1}{\omega_{0}^{2}-\omega^{2}-{\rm i}\gamma\omega}\left(\omega_{p0} - {\rm i}ha\frac{\partial}{\partial\omega}\right)\tilde{\boldsymbol{E}},\label{eq:P-sym}
\end{equation}
whereas Weyl quantization ($\alpha=\frac{1}{2}$) applied to Eq. (\ref{eq:pseudo-1}) yields,
\begin{align}
\tilde{\boldsymbol{P}} &= \epsilon_{0}\int_{-\infty}^{\infty} {\rm d}\tilde{\xi}_{1}\left[\left(\omega_{p0} - \frac{{\rm i}ha}{2}\left(\frac{\partial}{\partial\omega}+\frac{\partial}{\partial\tilde{\xi}_{2}}\right)\right)^{2}\frac{1}{\omega_{0}^{2}-\tilde{\xi}_{1}^{2}-{\rm i}\gamma\tilde{\xi}_{1}}\,\tilde{\boldsymbol{E}}\!\left(\tilde{\xi}_{1}/\Delta\right)\delta(\omega-\tilde{\xi}_{1})\right]_{\tilde{\xi}_{2}=\omega}\nonumber\\[5pt]
&= \epsilon_{0}\Bigg[\left(\omega_{p0} - {\rm i}ha\frac{\partial}{\partial\omega}\right)\frac{1}{\omega_{0}^{2}-\omega^{2}-{\rm i}\gamma\omega}\left(\omega_{p0} - {\rm i}ha\frac{\partial}{\partial\omega}\right)\tilde{\boldsymbol{E}}\nonumber \\[5pt]
&- \frac{h^{2}a^{2}}{4}\left(\frac{\partial^{2}}{\partial\omega^{2}}\!\frac{1}{\omega_{0}^{2}-\omega^{2}-i\gamma\omega}+ \frac{1}{\omega_{0}^{2}-\omega^{2}-i\gamma\omega}\frac{\partial^{2}}{\partial\omega^{2}}-2\frac{\partial}{\partial\omega}\frac{1}{\omega_{0}^{2}-\omega^{2}-i\gamma\omega}\frac{\partial}{\partial\omega}\right)\tilde{\boldsymbol{E}}\Bigg],
\end{align}
which contains additional terms compared to Eq. (\ref{eq:P-sym}).
\section{Details of Drude model permittivity used in semi-classical equations\label{app:drude-details}}
In Fig.~\ref{fig:semiclassical} the following permittivity was used
\begin{equation}
    \epsilon(\omega,\mathcal{T})=\epsilon_{\rm b}-\frac{\omega_{\rm p}^2(\mathcal{T})}{\omega(\omega+{\rm i}\gamma)}
\end{equation}
where $\gamma=0.25$ and $\epsilon_{\rm b}=3.7$, which corresponds to the wave-vector (operators replaced with numbers, as explained in Sec.~\ref{sec:wigner})
\begin{equation}
    K(\mathcal{T},\omega)=\omega\sqrt{\epsilon_{\rm b}-\frac{\omega_{\rm p}^2(\mathcal{T})}{\omega(\omega+{\rm i}\gamma)}}=\omega\sqrt{\epsilon_{\rm b}-\frac{\omega_{\rm p}^2(\mathcal{T})}{(\omega^2+\gamma^2)}+\frac{{\rm i}\gamma\omega_{\rm p}^2(\mathcal{T})}{\omega(\omega^2+\gamma^2)}}.
\end{equation}
For the particular case shown in Fig. \ref{fig:semiclassical} we took the plasma frequency to evolve as
\begin{equation}
    \omega_{\rm p}^2(t)=\omega_{\rm p,0}^2 - \delta\omega_{p}^2\left[1+\tanh\left(\frac{t-t_0}{\Delta_1}\right)\exp\left(-\frac{t-t_0}{\Delta_2}\right)\right]
\end{equation}
where $\omega_{\rm p,0}^2=96.0$ and $\delta\omega_{\rm p}^2=2.0$, $\Delta_1=3.5$, and $\Delta_{2}=45.0$.
\end{document}